\documentclass[namedreferences]{solarphysics}

\usepackage[hyperref,optionalrh]{spr-sola-addons} 
\usepackage{graphicx}        
\usepackage{color}           
\usepackage{breakurl}        

\newcommand{\aap}{    {\it Astron. Astrophys.}}

\newcommand{\apj}{    {\it Astrophys. J.}}
\newcommand{\apjl}{   {\it Astrophys. J. Lett.}}

\newcommand{\grl}{    {\it Geophys. Res. Lett.}}

\newcommand{\jgr}{    {\it J. Geophys. Res.}}

\newcommand{\solphys}{{\it Solar Phys.}}
 
\newcommand{\ssr}{    {\it Space Sci. Rev.}}

\chardef\us=`\_

\begin{document}

\begin{article}
\begin{opening}

\title{A Transient Coronal Sigmoid in Active Region NOAA 11909: Build-up Phase, M-class Eruptive Flare, and Associated Fast Coronal Mass Ejection}

\author[addressref=aff1,corref,email={hema@prl.res.in}]{\inits{Hema}\fnm{Hema}~\lnm{Kharayat}\orcid{0000-0003-3522-3135}}
\author[addressref=aff1,email={}]{\inits{B.}\fnm{Bhuwan}~\lnm{Joshi}\orcid{0000-0001-5042-2170}}\sep
\author[addressref=aff1,email={}]{\inits{p.}\fnm{Prabir K.}~\lnm{Mitra}\orcid{0000-0002-0341-7886}}
\author[addressref={aff2,aff3},email={}]{\inits{p.}\fnm{P. K.}~\lnm{Manoharan}\orcid{0000-0003-4274-211X}}
\author[addressref=aff4,email={}]{\inits{p.}\fnm{Christian}~\lnm{Monstein}\orcid{0000-0002-3178-363X}}
 


\address[id=aff1]{Udaipur Solar Observatory, Physical Research Laboratory, Udaipur, 313 001, India}
 \address[id=aff2]{Radio Astronomy Centre, National Centre for Radio Astrophysics, Tata Institute of Fundamental Research, Udhagamandalam (Ooty), 643 001, India}
 
 \address[id=aff3]{Arecibo Observatory, University of Central Florida, Puerto Rico, USA.}
 
\address[id=aff4]{Istituto Ricerche Solari Locarno (IRSOL), Via Patocchi 57, 6605 Locarno Monti, Switzerland}

\runningauthor{H. Kharayat et al.}
\runningtitle{Build-up and Disruption of a Transient Sigmoid}

\begin{abstract}
In this article, we investigate the formation and disruption of a coronal sigmoid from the active region (AR) NOAA 11909 on 07 December 2013, by analyzing multi-wavelength and multi-instrument observations. Our analysis suggests that the formation of the sigmoid initiated $\approx$ 1 hour before its eruption through a coupling between two twisted coronal loop systems. This sigmoid can be well regarded as of `transient' class due to its short lifetime as the eruptive activities started just after $\approx$ 20 min of its formation. A comparison between coronal and photospheric images suggests that the coronal sigmoid was formed over a simple $\beta$-type AR which also possessed dispersed magnetic field structure in the photosphere. The line-of-sight photospheric magnetograms also reveal moving magentic features, small-scale flux cancellation events near the polarity inversion line, and overall flux cancellation during the extended pre-eruption phase which suggest the role of tether-cutting reconnection toward the build-up of the flux rope.  The disruption of the sigmoid proceeded with a two-ribbon eruptive M1.2 flare (SOL2013-12-07T07:29). In radio frequencies, we observe type III and type II bursts in meter wavelengths during the impulsive phase of the flare. The successful eruption of the flux rope leads to a fast coronal mass ejection (with a linear speed of $\approx$ 1085 km s$^{-1}$) in SOHO/LASCO field-of-view. During the evolution of the flare, we clearly observe typical ``sigmoid-to-arcade" transformation.  Prior to the onset of the impulsive phase of the flare, flux rope undergoes a slow rise ($\approx$ 15 km s$^{-1}$) which subsequently transitions into a fast eruption ($\approx$ 110 km s$^{-1}$). The two-phase evolution of the flux rope shows temporal associations with the soft X-ray precursor and impulsive phase emissions of the M-class flare, respectively, thus pointing toward a feedback relationship between magnetic reconnection and early CME dynamics.

\end{abstract}
\end{opening}

\section{Introduction}
     \label{S-Introduction}

Solar flares are characterized by the sudden enhancement of localised brightness in the solar atmosphere. During a flare, a tremendous amount of energy, ranging as high as 10$^{28}$- 10$^{32}$ erg, is released which manifests its signatures in the entire electromagnetic spectrum \textit{i.e.} $\gamma$-rays to radio waves \citep[\textit{e.g.} see review by][]{Shibata2011}. It is now well understood that a flare may be accompanied by the expulsion of large amount of plasma from the corona into the interplanetary space, \textit{i.e.} Coronal Mass Ejection (CME). Contrary to such `eruptive flares', a second class of flares known as `confined flares', do not show any association with CMEs. The confined flares mostly undergo evolution in compact loop networks without causing significant reconfiguration of the large scale magnetic configuration \citep[see \textit{e.g.}][]{Upendra2014ApJ}. On the other hand, the eruptive flares and associated CMEs cause large-scale changes in the coronal magnetic structure. Further, Earth-directed CMEs are known for their geo-effectiveness. It is, therefore, essential to study the source region characteristics of CMEs for a better understanding of space weather phenomena.

Observations show that the CME-productive active regions exhibit some interesting features during the pre-eruption phase such as coronal sigmoids, filaments (or prominences), filament channels, and extreme ultraviolet (EUV) hot channels \citep[see reviews by][]{Gibson2006SSR,Alexander2006SSRv,Chen2011LRSP,Toriumi2019LRSP}. Among these features, filaments are the most studied structures associated with solar eruptions. Filaments are dark thread-like structures which consist of cool, dense plasma material embedded in the hot tenuous solar corona \citep{Martin1998}. These structures are traditionally observed in ground based H$\alpha$ images. When observed above the solar limb, these structures appear bright against dark background, hence, are termed as prominences \citep{Parenti2014}. Filaments exist along the magnetic neutral line or polarity inversion line with their legs rooted in the opposite polarity magnetic flux regions. The narrow lanes between opposite polarity magnetic fields where a filament can be formed, are called filament channels \citep{Engvold1997}. Space based observations have revealed important hot-plasma structures that form in the active region corona, such as, sigmoids and hot channels. Sigmoids are S (or inverse S) shaped structures which are observed in soft X-ray \citep[SXR;][]{Manoharan1996, Rust1996} and EUV emissions \citep{Cheng2014,Bhuwan2017}. With the advent of multi-channel EUV imaging from \textit{Atmospheric Imaging Assembly} \citep[AIA;][]{Lemen2012} on board \textit{Solar Dynamics Observatory} \citep[SDO;][]{Pesnell2012}, hot cohrent plasma structures (often termed as hot channels) have been identified from several CME producing active regions \citep{Cheng2013ApJ, Patsourakos2013ApJ, Nindos2015ApJ}. These features essentially provide evidence of the magnetic flux rope: a set of magnetic field lines winding around a common axis \citep{Gibson2006}. Contemporary, multiwavelength studies suggest these hot channels to be the earliest signature of a CME in the source region \citep{Cheng2011ApJ, Bhuwan2017,James2017SoPh, Prabir2019ApJ, Suraj2020}.

The formation mechanisms of the flux ropes in the corona have been addressed in many studies. It has been suggested that the presence of the coronal flux ropes is intrinsically related to their emergence from  subphotospheric layers \citep{Amari2004ApJ, Archontis2004AA}. Further, it has been recognised that a magnetic flux rope can be formed above the photosphere by shearing motions, emergence or submergence of the photospheric flux, or by the combinations of these processes; the above mechanisms essentially rely on the flux cancellation at the photosphere \citep{Kusano2005ApJ,Mackay2006ApJ,Green2009ApJ,Green2011AA,Cheng2013ApJ}. In view of the onset of an eruptive flare, the observations suggest that the flux rope associated structures can be formed either during a solar eruption or may exist well before it \citep{Zhang2012, Patsourakos2013ApJ, James2017SoPh, Prabir2020SoPh}. The studies further indicates that the \textit{in-situ} build-up of the flux rope during or just before an eruptive event is accomplished \textit{via} magnetic reconnections \citep{Cheng2011ApJ, Patsourakos2013ApJ, Kumar2016ApJ}. It is noteworthy that the existance of the flux rope in CMEs have been evidenced by near-Earth spacecraft measurements, in particular, by the in-situ detection of magnetic clouds \citep{Demoulin2016SoPh, Wang2018JGRA, Syed2019SoPh}. Although the origin and basic structures of the flux ropes have been recognised and studied in many previous studies, there is a lot more to learn about the formation process of the flux ropes and temporal aspects of their evolution with respect to the onset of CMEs in the source region. Simultaneous multi-channel observations obtained from Helioseismic Magnetic Imager \citep[HMI;][]{Schou2012} and AIA onboard SDO at high spatial and temporal resolutions have provided us with a unique opportunity for such investigations. The multi-wavelength observations ensure the examination of various layers of the solar atmosphere right from the photosphere to the lower corona. Once a flux rope formed in the corona, the dips in the helical magnetic structure provide support for the filament material against gravity \citep[][see also review by \citealp{Gibson2018LRSP}]{Kuperus1974AA, Gibson2006}. However, the presence of dipped magnetic structure does not provide the necessity for the filament. The existance of the filament channel is more frequent than the filament itself, pointing toward a magnetic environment which is able to support the filament even if the filament does not exists itself. The filament is often overlaid by the sigmoidal structure \citep{Gibson2002}. The relationship between the filament and the sigmoid have been discussed in many previous studies \citep[][see also review by \citealp{Gibson2018LRSP}]{Gibson2000JGR, Pevtsov2002SoPh, Regner2004AA,Bhuwan2017}. By using Yohkoh SXR data and H$_\alpha$ full disk observation, \cite{Pevtsov2002SoPh} have studied sigmoid-filament association for six active regions. He found a cospatial association between filament and sigmoid, through which he concluded that both features are related to same topological structure. Based on three-dimensional non-linear force free field modelling, \cite{Regner2004AA} have investigated the relationship between filament and sigmoid in the active region 8151. They have addresed that both the filament and the sigmoid can be described by long twisted flux tubes.

Sigmoids indicate the presence of sheared and weakly twisted coronal field lines with twist of around 1 turn in the field. These sheared and twisted magnetic field structures can store free magnetic energy \citep{Gibson2006SSR} and have been regarded as the source region of powerful CMEs. Using SXR images from the Soft X-ray Telescope (SXT) on the Yohkoh satellite, \cite{Hudson1998} examined the coronal structures associated with the halo CMEs and found a characteristic pattern of sigmoid-to-arcade development with the passage of a CME. Using SXT data, \cite{Canfield1999} classified active regions according to their morphology (sigmoidal or non-sigmoidal) and nature of activity (eruptive or non-eruptive). They have found that the sigmoidal active regions are prone to be eruptive than the non-sigmoidal regions. In the case study of a major geo-effective CME from acive region NOAA 12371, \cite{Joshi2018SoPh} found CME to be associated with a coronal sigmoid that displayed an intense emission from its core before the onset of the eruption. Various studies have been performed in order to understand the formation and eruption of the sigmoids \citep{Sterling2000, Savcheva2012, Bhuwan2017}. \cite{Bhuwan2017} studied the formation of a coronal sigmoid in the active region NOAA 11719 and found that the development of sigmoid structure occurred through the successive interaction between two J-shaped bundles of loops. \cite{Prabir2018ApJ} examined two major X-class solar flares from the sigmoidal active region NOAA 12673 and found that the complex network of $\delta$-sunspots in the active region resulted to the formation of the sigmoid in the hot EUV channel. Their study also revealed the flux rope structure by coronal magnetic field modeling. The multiwavelength case studies of the build-up processes of the coronal sigmoids and early evolutionary phases of CMEs from these regions are therefore extremely important toward understanding the origin of large-scale solar eruptions and development of methods for their forecast.

In view of the above scientific motivations, here we present a comprehensive multi-wavelength analysis of the formation of coronal sigmoid and subsequent disruption from the region that resulted into a fast CME ($\approx$ 1085 km s$^{-1}$) and an M-class two-ribbon flare (SOL2013-12-07T07:29). The sigmoidal flux rope was developed over a magnetically weaker and dispersed bipolar active region (AR) NOAA 11909 on 07 December 2013. Further, sigmoidal structure disrupted just $\approx$ 20 min after its complete formation; hence, it can be justifiably called a `transient' sigmoid. The flux rope activates during the precursor phase of the M1.2 flare during which the flux rope slowly rose. Subsequently, the flux rope moved to the phase of eruptive expansion with the onset of impulsive flare emission and exhibited important observational signatures in EUV and radio wavelengths. The paper is organized as follows: in Section~\ref{method}, we discuss a brief account of the observational data and method used for the study.  Multi-wavelength observations and results are presented in Section~\ref{results}. We discuss and interpret the results in Section~\ref{discussion}. A summary of the results is provided in Section~\ref{conclusion}.

\section{Observational Data and Method} 
      \label{method}  
      
     For the multi-wavelength study of M1.2 flare on 07 December 2013 and its source active region (NOAA 11909), we analyze data from multiple ground and space borne instruments which are briefly discussed below:

\begin{enumerate}
	\item Observations of the different solar atmospheric layers in (E)UV wavelengths are obtained from the \textit{Atmospheric Imaging Assembly} \citep[AIA:][]{Lemen2012} on board \textit{Solar Dynamics Observatory} \citep[SDO:][]{Pesnell2012}. SDO/AIA provides solar images in seven EUV channels: 94 \AA, 131 \AA, 171 \AA, 193 \AA, 211 \AA, 304 \AA, and 335 \AA;~two UV channels: 1600 \AA~ and 1700 \AA; and one white light channel: 4500 \AA. AIA provides 4096 $\times$ 4096 pixel images at a high resolution of 0".6 per pixel with a cadence of 12 sec.  

\item For the photospheric observation of the Sun, we have used the data provided by the \textit{Helioseismic Magnetic Imager} \citep[HMI:][]{Schou2012} of SDO. HMI has spatial resolution of 0".5 per pixel. We used white light continnum intensity images and line-of-sight (LOS) magnetograms of the solar photosphere with a temporal cadence of 45 sec.

\item CME observations are analysed by the C2 and C3 white light coronagraphs of \textit{Large Angle and Spectrometric Coronagraph} \citep[LASCO:][]{Brueckner1995} on board the \textit{Solar and Heliospheric Observatory} \citep[SOHO:][]{Domingo1995}. The field-of-view (FOV) for C2 and C3 coronagraphs are 1.5--6 R$_\odot$ and 3.7--30 R$_\odot$, respectively.

\item To study radio emission associated with this eruptive flare, we have analyzed the data obtained from \textit{extended Compact Astronomical Low-cost Low-frequency Instrument for Spectroscopy and Transportable Observatory}\footnote{http://www.e-callisto.org/} \citep[e-CALLISTO;][]{Benz2005}. The radio observations presented in this study are taken from the Ooty station of the e-CALLISTO network.

\item We have used the H$\alpha$ observation with the Solar Flare Telescope \citep[SFT;][]{Ichimoto1991} operated at National Astronomical Observatory of Japan (NAOJ), Mitaka.  SFT is the ground-based instrument  of Solar Science Observatory.

\item Soft X-Ray (SXR) images of the Sun has been taken from the observations of \textit{Solar X-ray Imager} (SXI) aboard \textit{Geostationary Operational Environmental Satellite}  \citep[GOES;][]{Bornmann1996SPIE}. SXI provides the full-disk images of the Sun in X-rays with a temporal cadence of one minute.
\end{enumerate}

 In this study, coronal magnetic field lines are modeled by employing the potential field extrapolation method by solving a Green's function \citep{Seehafer1978SoPh} which is available in the SolarSoftware library encoded in the Interactive Data Language (IDL). As the boundary condition for potential field extrapolation, we used z-component of photospheric vector magnetogram of 06:15 UT from hmi.sharp\_cea\_720s series of HMI/SDO at a spatial resolution of 0".5 pixel$^{-1}$ and temporal cadence of 720s. Here, the HMI magnetogram has a dimension of 812$\times$462 pixels which corresponds to a size of $\approx$294$\times$160 Mm. Extrapolation has been performed up to a height of $\approx$160 Mm from the photosphere. The extrapolated field lines are visualised by using Visualization and Analysis Platform for Ocean, Atmosphere, and Solar Researchers \citep[VAPOR;][]{Clyne2007NJPh} software.

\section{Multiwavelength Observations and Results} 
  \label{results}
  
 \subsection{Overview of Flaring Active Region NOAA 11909}
 \label{S-event overview}
 
 On the day of the reported activity \textit{i.e.} 07 December 2013, the AR NOAA 11909 was situated at the heliographic coordinate S16W49. This AR emerged from the eastern limb of the Sun on 28 November 2013 as $\beta$-type sunspot and moved to the far side of the Sun over the western limb on 10 December 2013 as $\alpha$-type. During its lifetime, its magnetic configuration evolved as $\beta$ - $\beta\gamma$ - $\beta$ - $\alpha$- type and produced three C-class and only one M-class solar flares. 
 
  \begin{figure}    
 	\centerline{\includegraphics[width=.57\textwidth,clip=]{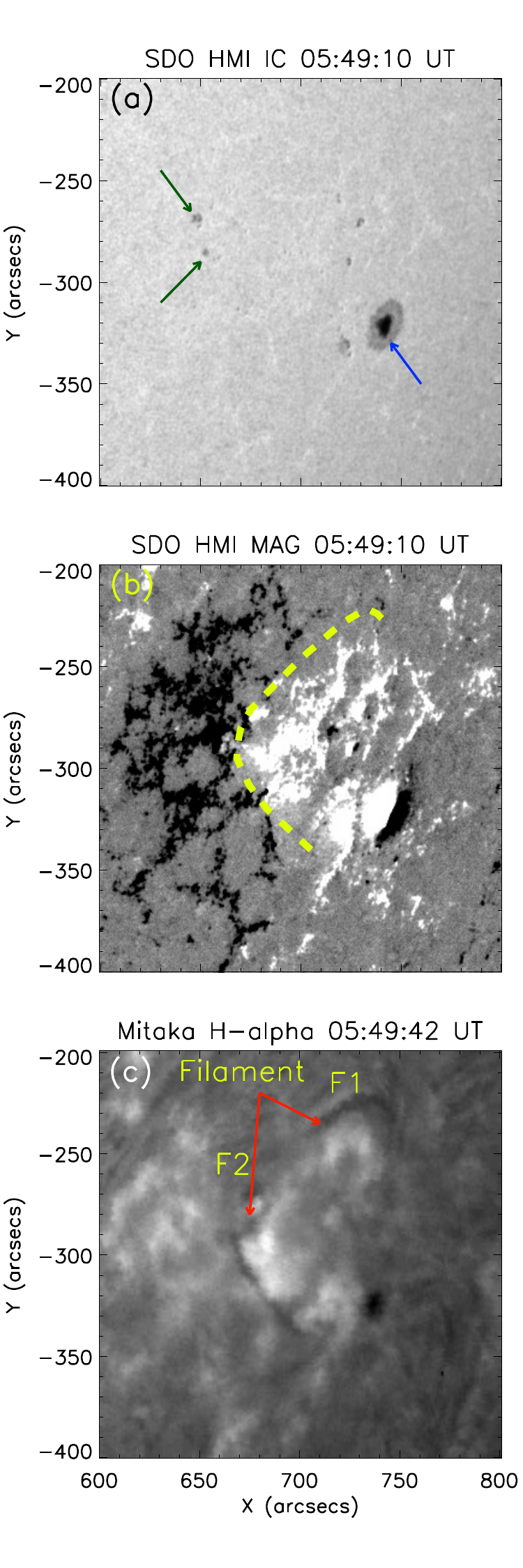}
 	}
 	\caption{Multi-wavelength view of AR NOAA 11909 on 07 December 2013 prior to the event, (a) HMI white light image of AR showing the distribution of sunspots. Small and large size sunspots are indicated by green and blue arrows, respectively. (b) Cotemporal HMI LOS magnetogram showing the magnetic field configuration of the AR 11909. Dashed yellow curve indicate the approximate polarity inversion line (c) Mitaka H-alpha image showing the presence of filaments F1 and F2 (marked by red arrows) at the studied AR.} 
 	\label{Fig1}
 \end{figure}
 
 Figure~\ref{Fig1} shows the multi-wavelength view of AR NOAA 11909 on 07 December 2013 at 05:49 UT, prior to the onset of the M-class flare. Figures~\ref{Fig1}a and b display the photospheric continnum intensity image and LOS magnetogram, respectively. It is clear from these panels that the AR consists of several small to medium size sunspots with dispersed magnetic field. Among these sunspots, the largest one (indicated by blue arrow) form the main sunspot of the leading group. Comparison of Figures~\ref{Fig1}a and b confirms that the leading sunspot group largly consists of positive magnetic polarity while the trailing part of the AR is of negative polarity. Notably, magnetic flux in the trailing sunspot group is quite dispersed and composed of a couple of smaller sunspots (shown by green arrows) along with small flux element of the plage region. The positive and negative flux regions exhibit a clear seperation so that a polarity inversion line (PIL) can be easily drawn which is delineated by a dashed yellow curve in Figure~\ref{Fig1}b. The clear distinction of bipolar region suggests this AR to be a $\beta$-type, prior to the reported activity. H$\alpha$ image of AR shows the presence of two filaments: F1 and F2, lying approximately along the PIL which are indicated by red arrows in Figure \ref{Fig1}c. Filament F1 lies in the northern part while filament F2 lies in the southern part of the PIL.
 
  \begin{figure}    
 	\centerline{\includegraphics[width=0.90\textwidth,clip=]{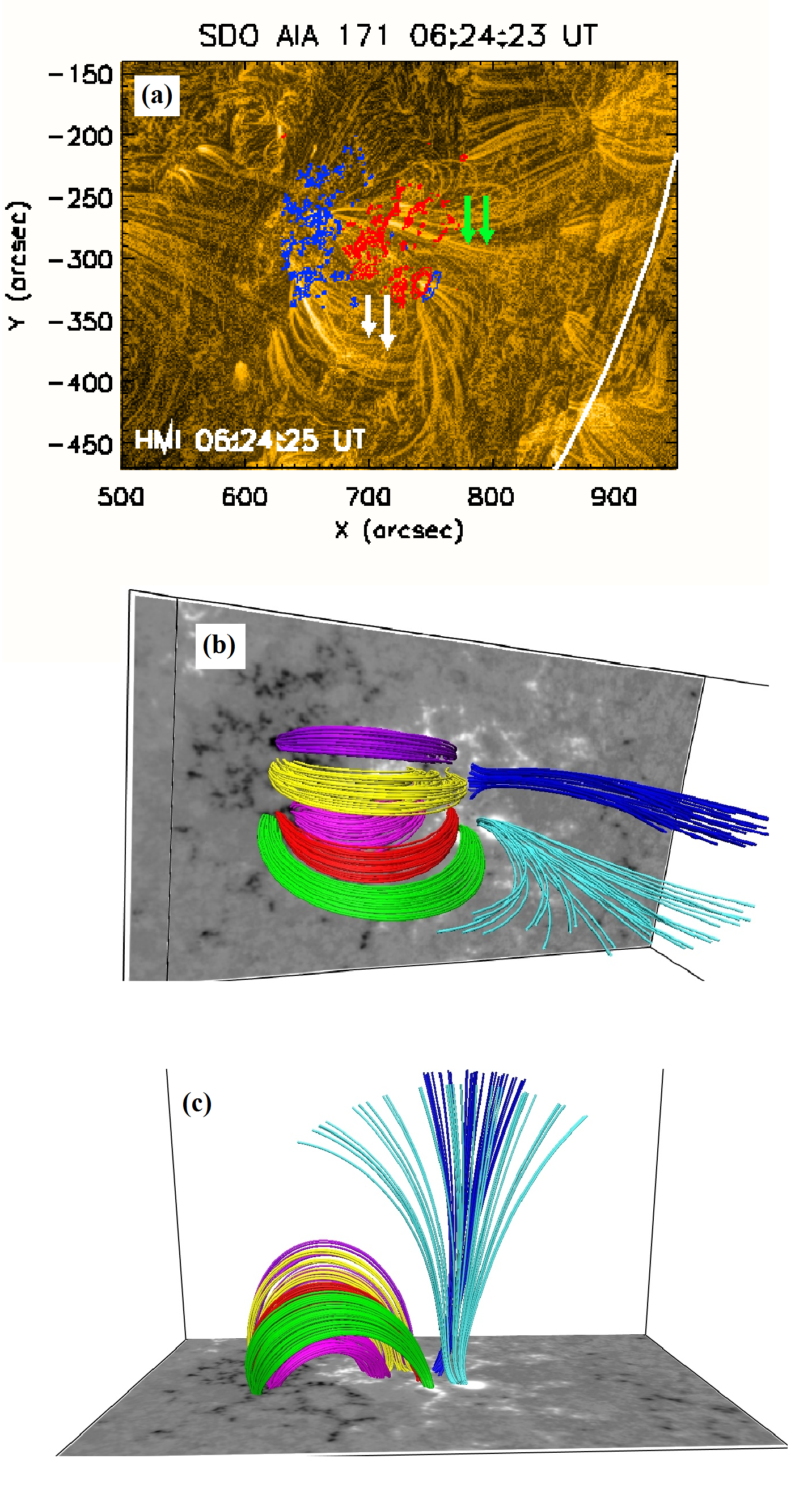}
 	}
 	\caption{(a) AIA EUV image of the AR 11909 in 171 \AA\ channel on 07 December 2013 at 06:24 UT processed by multi-scale gaussian normalisation technique. HMI line-of-sight magnetogram contours are overplotted to show the distribution of the photospheric flux. The positive and negative polarities of the flux are shown by red and blue contours, respectively. Contour levels are set as $\pm$150, $\pm$400, and $\pm$600 G. White and green arrows show the closed and open loops, respectively. (b-c) Modeled coronal magnetic field configuration based on the potential field extrapolation method. Different sets of closed and open field lines are shown by different colors, from front and side views, in panels (b) and (c), respectively.}
 	\label{Fig2}
 \end{figure}

 In Figure \ref{Fig2}a, we show AIA 171 \AA~image of the active region at 06:24 UT. For a clear visualisation of coronal features, the image is processed by Multi-Scale Gaussian Normalisation \citep[MGN;][]{Morgan2014SoPh} technique, which is based on localised normalisation of the data over a range of spatial scales. In order to display the distribution of the photospheric LOS magnetic field, we have overplotted HMI magnetogram as contours. Red and blue contours are used to distinguish positive and negative polarity fluxes, respectively. Two distinct sets of coronal loops can be observed from the AIA 171~\AA\ image of the AR (Figure~\ref{Fig2}a) that resemble a closed (white arrows) and an open (green arrows) magnetic field configurations. Notably, the following sunspot group is associated with closed field lines while closed as well as open field lines originate from the leading sunspot group. To confirm the presence of open field lines in the AR, we have extrapolated coronal magnetic field configuration by the use of potential field extrapolation method (see Section~\ref{method}). In Figures~\ref{Fig2}b and c, we display extrapolated coronal field lines from front and side views, respectively. We noticed the presence of different sets of closed coronal loops, connecting the positive flux region of leading sunspot group with the negative flux of the following sunspot group. The AR also contains open field lines which originate from the positive polarity regions of the leading sunspot group (shown by blue and cyan color lines).

\subsection{Build-up of Coronal Sigmoid in the Active Region} 
  \label{S-coronal sigmoid}
 
 \begin{figure}    
 	\centerline{\includegraphics[width=1.0\textwidth,clip=]{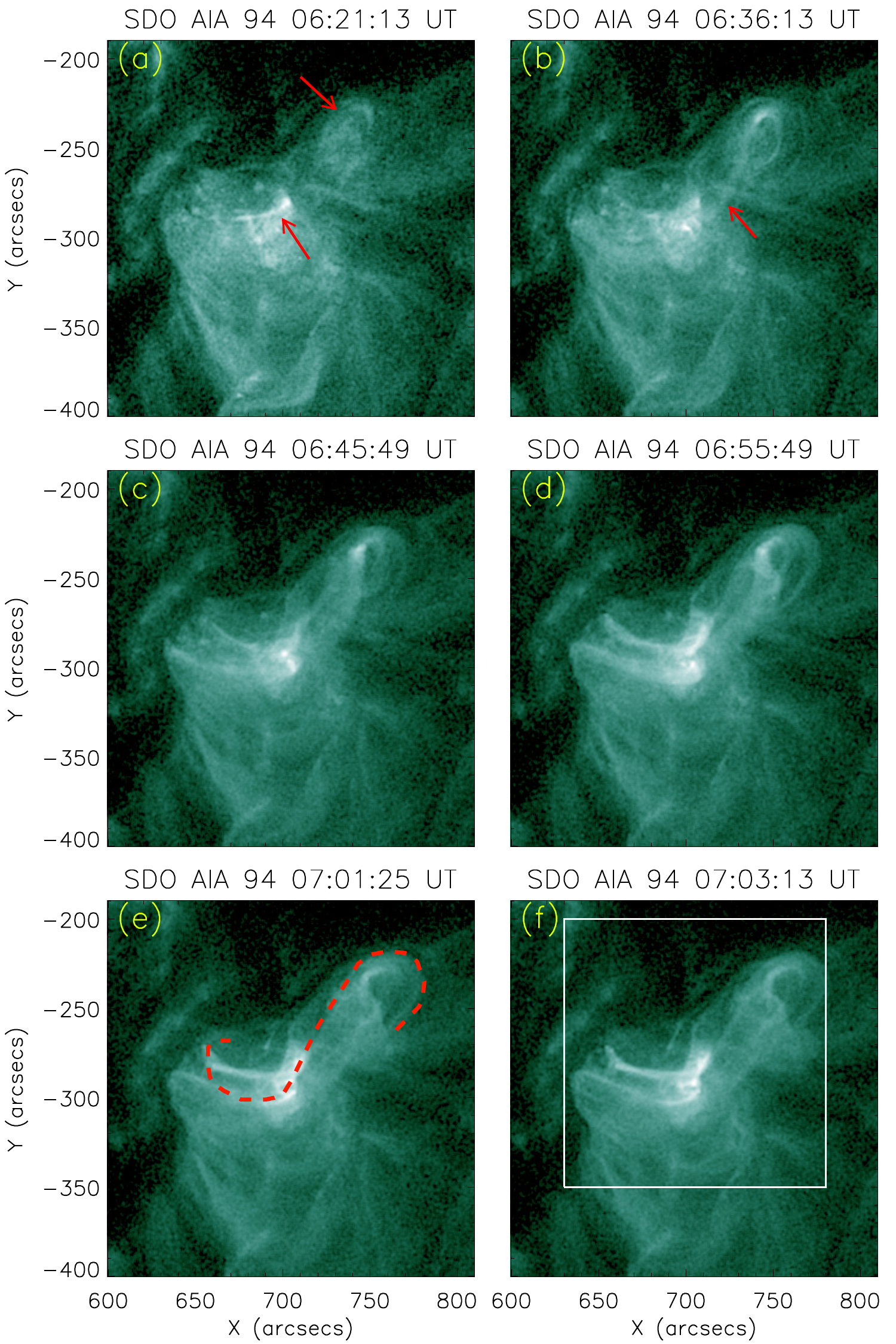}
 	}
 	\caption{Series of selected AIA EUV 94 {\AA} images showing the development of coronal sigmoid. Two arrows in panel (a) indicate two sets of coronal loops. Arrow in panel (b) marks the connectivity between two sets of loops. The dashed curve in panel (e) indicates the sigmoidal structure in the AR 11909. White box in panel (f) marks the region used to construct the AIA light curves which are plotted in Figure~\ref{Fig3}b.}
 	\label{Fig4}
 \end{figure} 

\begin{figure}    
	\centerline{\includegraphics[width=1.0\textwidth,clip=]{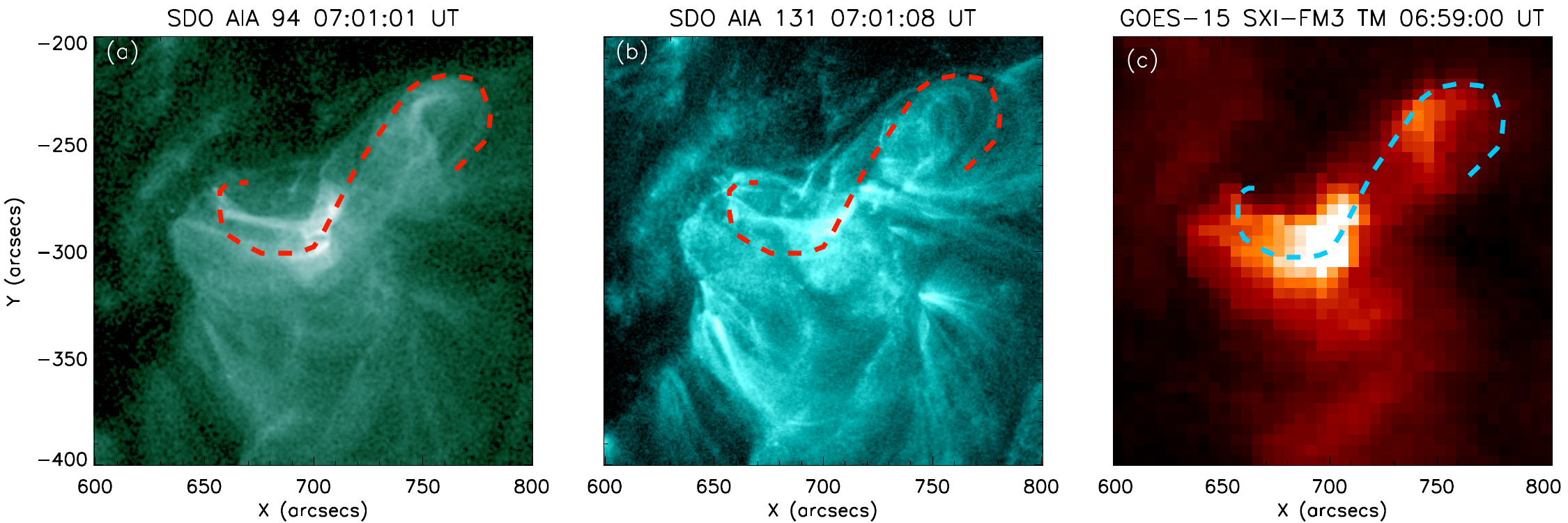}
	}
	\caption{An overview of the sigmoid in (a) AIA EUV 94~\AA, (b) 131~\AA, and (c) soft X-ray data from GOES-15. Dashed curve in all the panels marks the sigmoidal structure.}
	\label{Fig}
\end{figure}

\begin{figure}    
	\centerline{\includegraphics[width=1.0\textwidth,clip=]{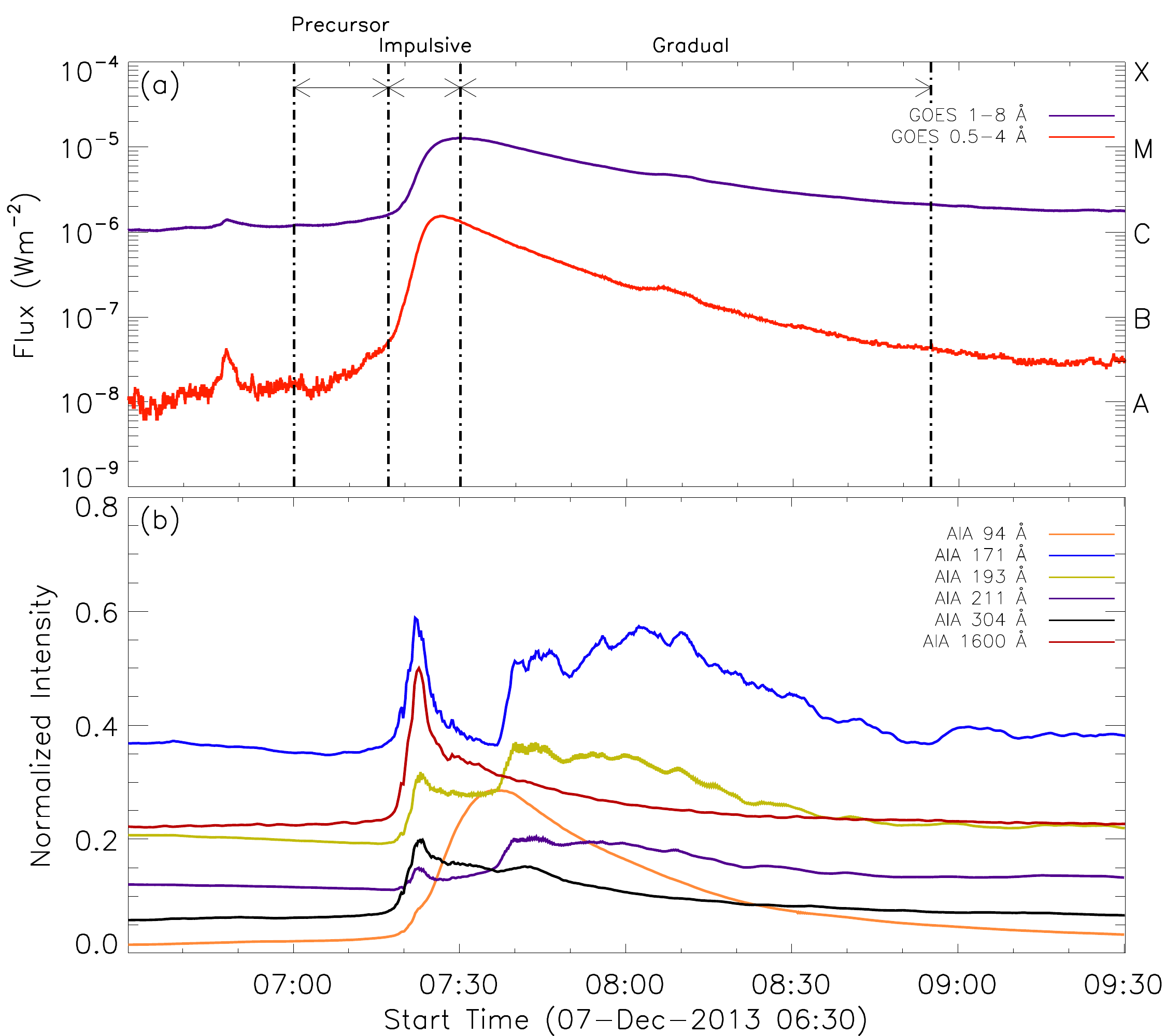}
	}
	\caption{(a) GOES SXR flux in 1--8 {\AA} and 0.5--4 {\AA} channels on 07 December 2013 from 06:30 UT to 09:30 UT. The vertical lines are used to differentiate the different phases of the M1.2 flare. (b) Cotemporal AIA light curves in 94 \AA, 171 \AA, 193 \AA, 211 \AA, 304 \AA, and 1600 {\AA} channels, normalized by their corresponding peak fluxes. In order to get clear visualisation, AIA light curves are further normalized by factor 3.5, 1.7, 2.7, 4.9, 5, and 2 for 94 \AA, 171 \AA, 193 \AA, 211 \AA, 304 \AA, and 1600 \AA, respectively. The region used to construct the AIA light curves is shown by the white box in Figure~\ref{Fig4}f.}
	\label{Fig3}
\end{figure}

 During $\approx$ 06:20 UT--07:00 UT, we observed signatures of the development of a coronal sigmoid. In Figure~\ref{Fig4}, we present a series of AIA 94 \AA~images  which show sequential formation of sigmoid in the activity region. Much before the onset of the flare ($\approx$ 1 hour), two bundles of coronal loops are identified which are indicated by the arrows in Figure \ref{Fig4}a. At this stage, we cannot identify any connectivity between these two bundles of loops. From $\approx$ 06:36 UT, a connection starts to develop between the loops (shown by arrow in Figure \ref{Fig4}b). From  $\approx$ 06:36 UT--06:55 UT, the connecting loop system undergoes through continuous evolution and expansion. As a result, we observe the establishment of the sigmoid at $\approx$ 07:00 UT (shown by red curve in Figure \ref{Fig4}e) which provides an evidence for the presence of the magnetic flux rope. A multi-wavelength view of the  fully developed sigmoid is presented in Figure~\ref{Fig}. Figures~\ref{Fig}a and b display the sigmoid in EUV AIA 94 \AA\ and 131 \AA\ channels, respectively. Figure~\ref{Fig}c shows the SXR image of the sigmoid. For a clear identification of the sigmoid, we have marked the sigmoid structure by a dotted curve in all of the panels. This sigmoidal flux rope appeared till the begining of the M-class flare at $\approx$ 07:17 UT. Afterward, the sigmoid disrupted as the eruption of the flux rope proceded (described in Section \ref{S-fluxrope}).

\subsection{Evolutionary Phases of M1.2 Flare} 
\label{phases}
Figure~\ref{Fig3}a depicts the temporal variation of the SXR flux in the 1--8 \AA\ (shown by violet curve) and 0.5--4 \AA\ (shown by red curve) wavelength bands of the GOES during the time interval 06:30 UT--09:30 UT on 07 December 2013. The evolution of the flare can be summarized by three phases: precursor ($\approx$07:00 UT--07:17 UT), impulsive ($\approx$07:17 UT--07:30 UT), and gradual phase ($\approx$07:30 UT--08:55 UT). The precursor phase is characterised by small-scale and gradual energy release during the slow rise of the sigmoidal flux rope. For characterising the impulsive phase of the flare, we have analyzed the temporal variation of the X-ray flux in the available hardest X-ray energy band \textit{i.e.} 3.0--25.0 keV which corresponds to the 0.5--4~\AA~wavelength band of GOES. A sharp increase in the X-ray flux at $\approx$ 07:17 UT indicates the start of the impulsive phase of the flare. A sudden enhancement in the intensity in most of the EUV channels at $\approx$ 07:17 UT (discussed later in Figure~\ref{Fig3}b), also signifies the start of the impulsive phase. The flare reached its peak intensity at $\approx$ 07:29 UT after which X-ray flux decrease gradually, defining the gradual phase of the flare. The evolutionary phases of the flare are described in Table~\ref{Tab:1}. The temporal evolutions of (E)UV intensities in the  94 \AA~(6 MK), 171 \AA~(600,000 K), 193 \AA~(1 MK), 211 \AA~(2 MK), 304 \AA~(50,000 K), and 1600 \AA~(10,000 K) wavelength bands of SDO/AIA are presented in Figure~\ref{Fig3}b. The region selected for constructing the AIA light curves is shown by the white box in Figure~\ref{Fig4}f. This region has been chosen because the flaring activity during the studied period was occurred within this region. It is to be noted that the intensity variation at the hot AIA 94 \AA\ channel resembles the GOES soft X-ray profiles with some time delay in their peak flux which indicate that, in spite of different temperature sensitivities of the instruments, both emissions originate from hot flaring plasma filled in the coronal loops. Interestingly, time profiles of the intensity variations in other AIA channels (except 1600~\AA) display two stage evolution. In the first stage, a sudden rise and fall of the intensity is observed while in the second stage the intensity variation is rather gradual. The intensity variation in AIA 1600~\AA~channel shows only single stage evolution which resembles the first stage of other AIA channels, \textit{viz.} 171 \AA, 193 \AA, 211 \AA, and 304 \AA. Notably, the first stage of the intensity variation in these channels is noticed earlier than the peak of GOES flux.

\begin{table}[]
	\caption{Summary of evolutionary phases of the M1.2 flare.}  \label{Tab:1}
	
	\begin{tabular}{cccl}
		\hline
		Sr.No. & Phase & Duration & Remarks  \\
		\hline
		1.& Build-up of & $\approx$ 06:20 UT--07:00 UT & Two twisted coronal loop systems    \\
		& coronal sigmoid&& are observed that sequentially join   \\
		&&& and form the coronal sigmoid.  \\
		2.& Precursor  &  $\approx$ 07:00 UT--07:17 UT & The flux rope undergoes  a slow rise; \\
		&&&  localized brightening suggesting \\
		&&& small-scale magnetic reconnections.\\
		3.& Impulsive & $\approx$ 07:17 UT--07:30 UT & The flux rope undergoes eruptive   \\ 
		&&& expansion; development of two  \\
		&&& parallel ribbons; type III and type \\
		&&& II radio bursts; ``sigmoid-to-arcade" \\
		&&&transformation begins.\\
		4.& Gradual & $\approx$ 07:30 UT--08:55 UT & A fast CME is first detected by  \\
		&&&LASCO C2 at $\approx$ 07:36 UT; highly\\
		&&&  structured post-flare loop arcade is \\
		&&& formed; ``sigmoid-to-arcade" \\
		&&&transformation completes.\\
		\hline
	\end{tabular}
\end{table}

\subsection{Photospheric Magnetic Field Variations} 
\label{flux:variation}

\begin{figure}[H]    
	\centerline{\includegraphics[width=1.0\textwidth,clip=]{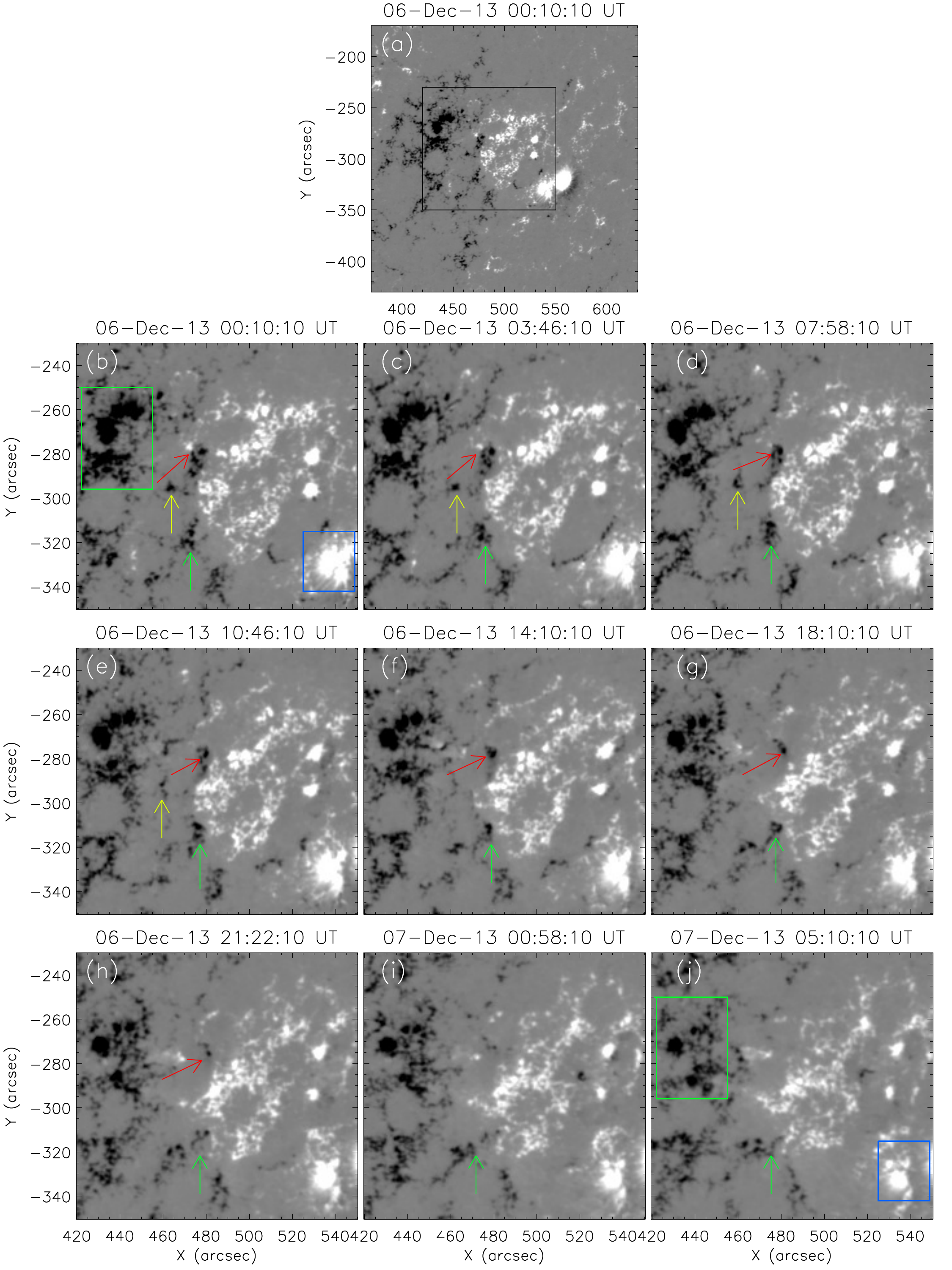}
	}
	\caption{(a) HMI magnetogram of the AR NOAA 11909 at 00:10 UT on 06 December 2013. Black box marks the region where we have observed the variations in the magnetic flux. (b-j) Zoomed-in view of black box region showing the changes in the magnetic flux. Green and blue boxes indicate the regions of large variation in negative and positive polarity fluxes, respectively. Different color arrows indicate different changing elements of magnetic flux near the PIL. An animation of this figure is provided in the supplementary materials.}
	\label{Fig5}
\end{figure}

In order to understand the formation of sigmoid and triggering mechanism of the eruptive flare, it is important to examine the photospheric magnetic field changes in the AR. Such changes result in the storage of magnetic energy in the corona and start to contribute much before the eruption. In Figure~\ref{Fig5}a we provide HMI magnetogram on 06 December 2013 at 00:10 UT, one day prior to the flaring activity. The region with significant photospheric changes is marked by the black box in Figure \ref{Fig5}a (see the animation associated with Figure~\ref{Fig5}). To probe, further magnetic changes, we provide a series of selected magnetograms of the marked subregion from 00:10 UT on 06 December 2013 to 05:10 UT on 07 December 2013. The green and blue boxes in Figures \ref{Fig5}b and \ref{Fig5}j show the regions associated with the prominent changes in negative and positive polarity fluxes, respectively. Comparing these two panels, we can clearly observe a significant cancellation of the negative polarity flux as well as decay of the positive polarity flux.  A small element of positive polarity is noticeable near the patch of negative polarity (indicated by red arrow in Figure~\ref{Fig5}b). The negative polarity patch undergoes continuous movement in northward direction and cancels out the positive polarity element (marked by red arrow in Figure~\ref{Fig5}d). The negative polarity patch continues its moving behaviour and disappears in the subsequent intervel (marked by red arrows in Figure~\ref{Fig5}e--h). Few other moving magnetic features of negative polarity are also noted near the PIL (indicated by yellow and green arrows in Figures~\ref{Fig5}b--j) which also get cancelled out subsequently.

\begin{figure}[H]    
	\centerline{\includegraphics[width=1.0\textwidth,clip=]{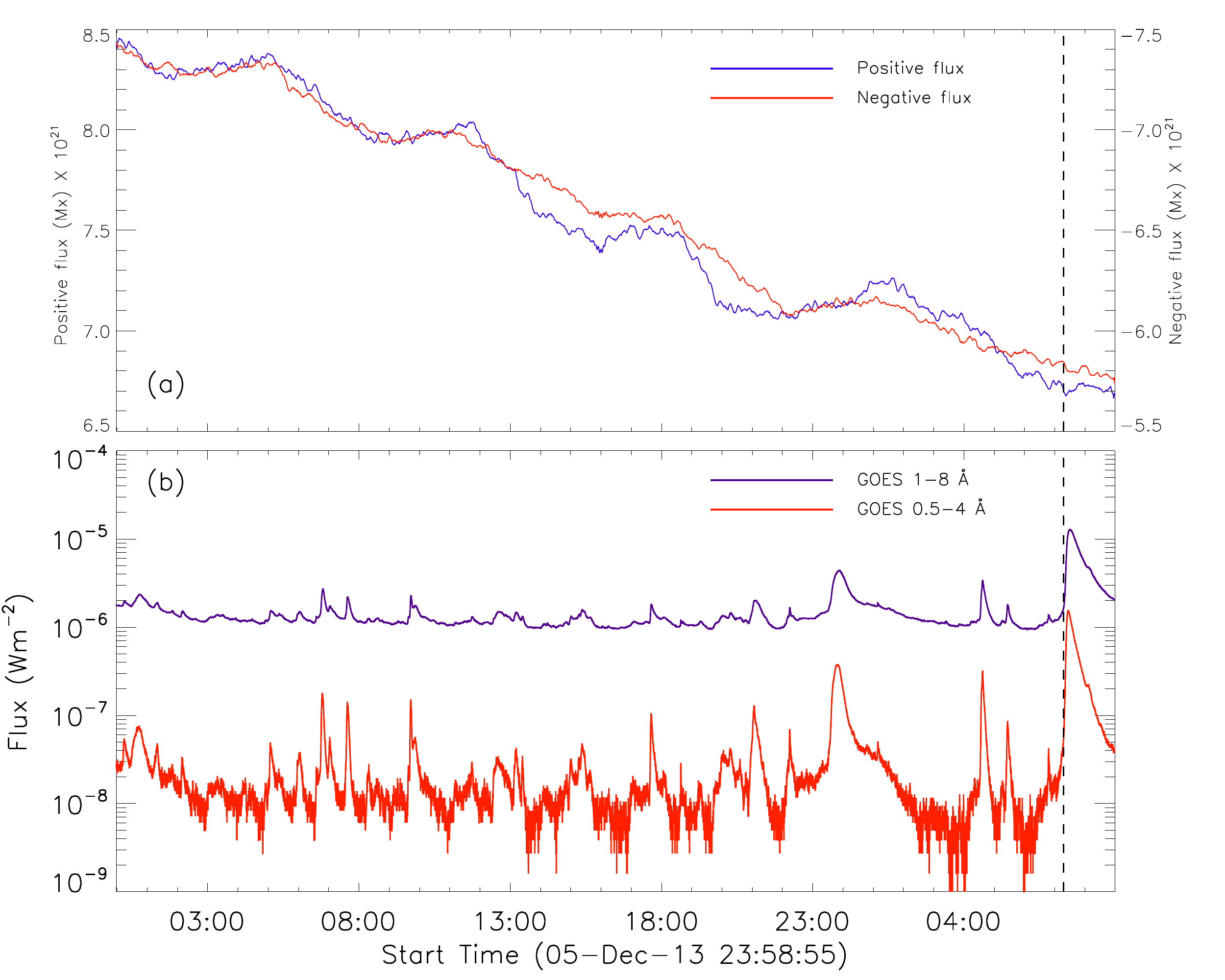}
	}
	\caption{(a) The temporal change of the photospheric magnetic flux of positive and negative polarities. (b) Cotemporal GOES SXR flux variation in 1--8 \AA\ and 0.5--4 \AA\ channels. The vertical dashed line in both the panels mark the onset of the M1.2 flare.}
	\label{Fig6}
\end{figure}

In Figure~\ref{Fig6}a, we plot the positive and negative fluxes through the whole AR (as shown in Figure~\ref{Fig5}a). For comparing magnetic flux variation with the coronal energy release, we plot the GOES SXR flux in 1--8 and 0.5--4 \AA\ wavelenth channel\textbf{s} in Figure~\ref{Fig6}b. We have found that the positive and negative polarity magnetic fluxes show large variations before the onset of the flaring activity. It decreases in a periodic manner. About five hour prior to the event (at $\approx$ 02:00 UT on 07 December 2013), the positive and negative fluxes start to decrease continuously. With the onset of the M1.2 flare (indicated by vertical dashed line in Figures \ref{Fig6}a and b), both the fluxes maintain an almost steady level. 

\subsection{Eruption of Sigmoidal Flux Rope} 
\label{S-fluxrope}

\begin{figure}[H]    
	\centerline{\includegraphics[width=1.0\textwidth,clip=]{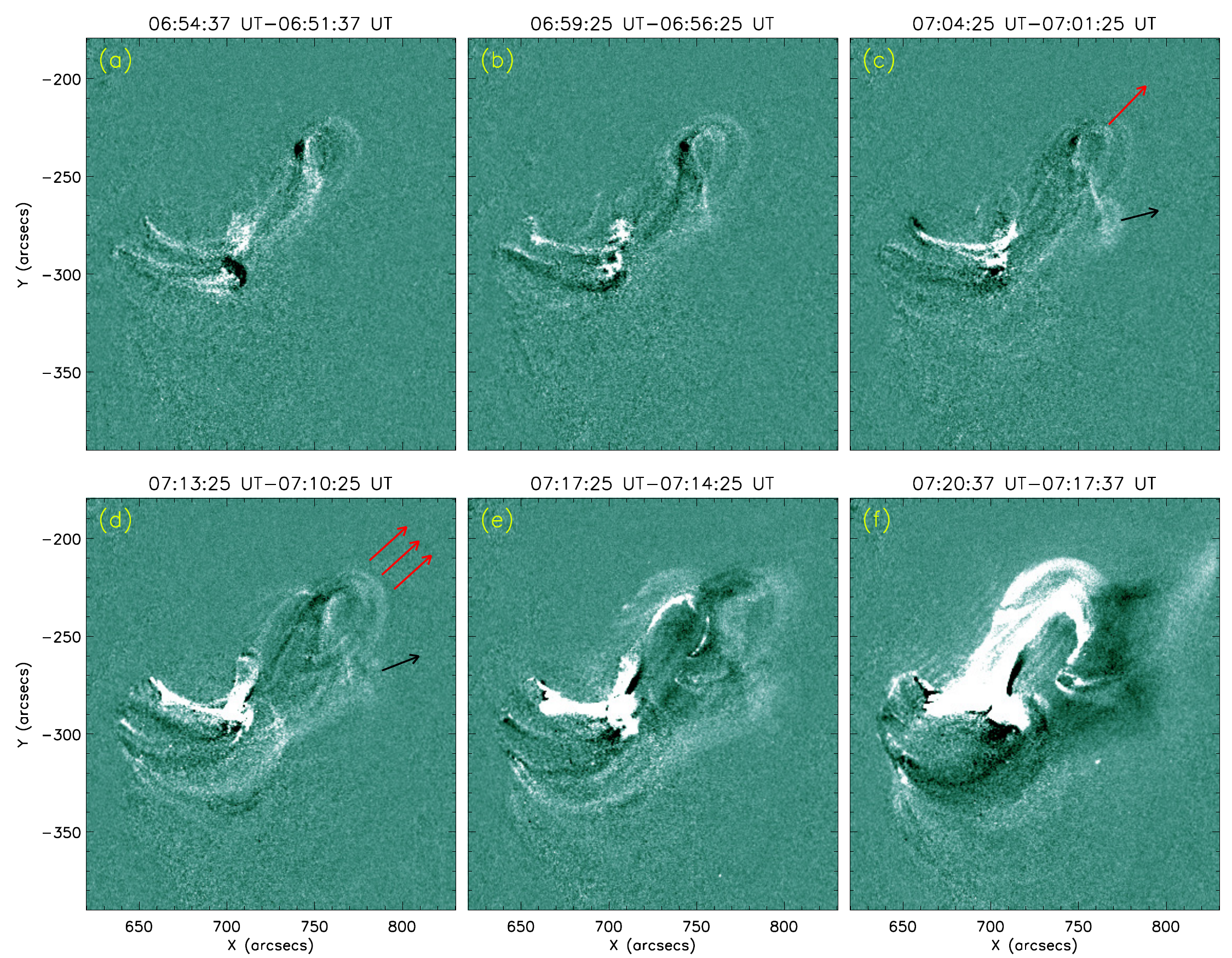}
	}
	\caption{SDO/AIA 94 \AA~running difference images showing the eruption of the flux rope between $\approx$ 06:54 UT and $\approx$ 07:20 UT. Two arrows in panel (c) show the two erupting fronts of the sigmoid. Arrows in panel (d) show the eruption of the sigmoid fronts.}
	\label{Fig9}
\end{figure}

Examination of series of AIA 94 \AA\ images suggests expansion of the sigmoidal flux rope structure after $\approx$ 07:00 UT which subsequently proceeds with its complete eruption. For a clear visualisation of the sigmoidal flux rope eruption, a series of AIA 94 \AA~running difference images are presented in Figure~\ref{Fig9}. It is observed that the eruption of the flux rope starts in the precursor phase at $\approx$ 07:03 UT in the northwest direction with two erupting fronts, which we call as the leading front (red arrows in Figures \ref{Fig9}c and d) and the following front (black arrows in Figures \ref{Fig9}c and d). The eruption of the leading front is clearly visible for an extended time interval encompassing the precursor and impulsive phases of the flare. However, the signatures of expanding following front are hard to distinguish during the impulsive phase and subsequently diminishes. The leading edge of the flux rope is shown by red arrows in Figure~\ref{Fig9}d. Coincidently, the leading front of the erupting flux rope is associated with the ``elbow" of the sigmoidal structure. Unlike the leading edge, the eruption of the central region of the sigmoidal structure is much less pronounced during the precursor and main phases of the flare. Hence, to understand the kinematic evolution of the flux rope at the source region we consider the height-time measurements of the leading front.

\begin{figure}[H]    
	\centerline{\includegraphics[width=1.0\textwidth,clip=]{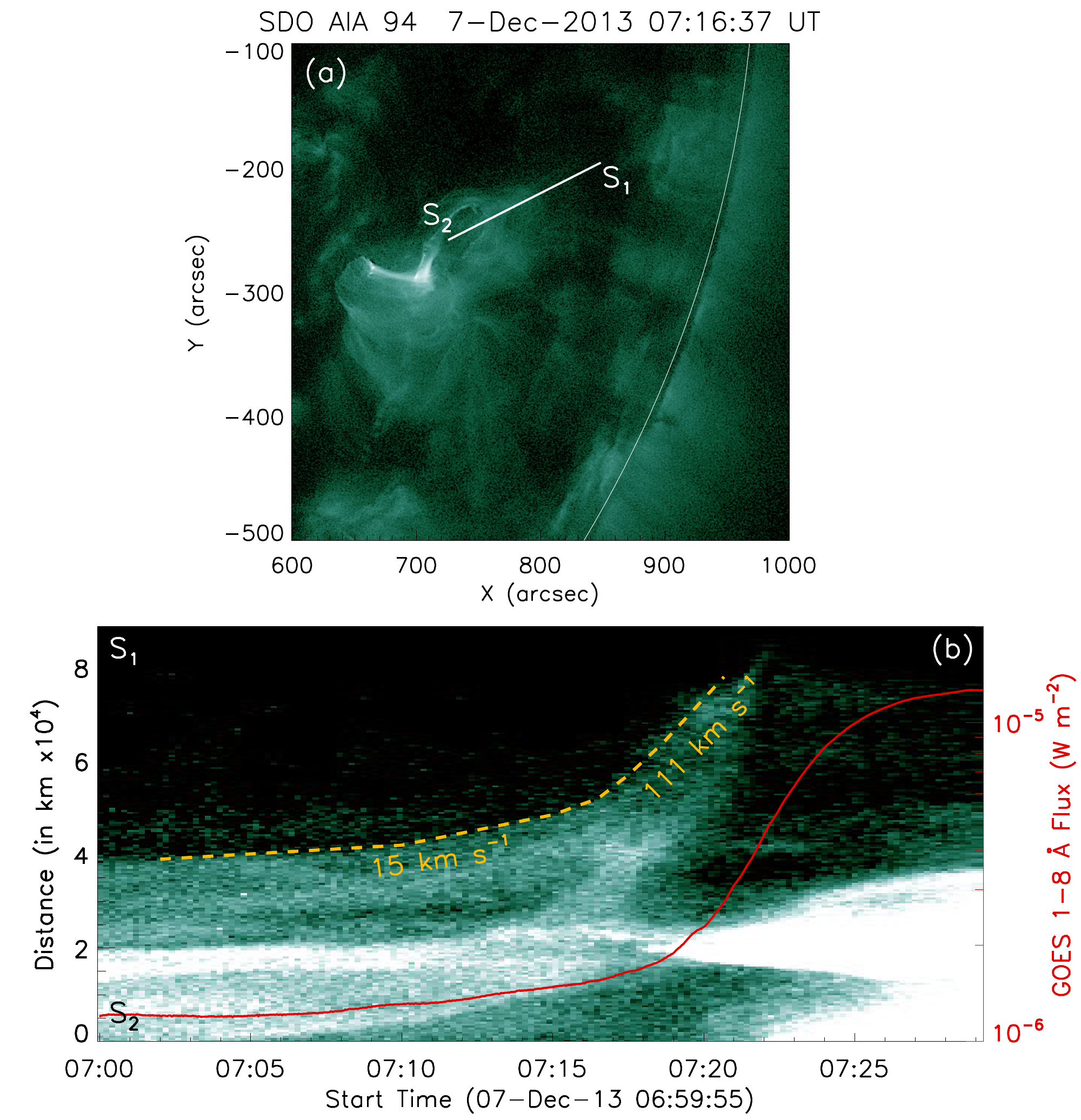}
	}
	\caption{(a) An AIA 94~\AA~image on 07 December 2013 at 07:16 UT with a straight line S$_2$S$_1$ along which  we have constructed a time-slice diagram. (b) Time-slice diagram showing the eruption of the flux rope during 07:00 UT to 07:29 UT. Red curve shows the temporal variation of the GOES SXR flux for the same time interval. In the precursor phase, we have observed a slow rise of the flux rope with linear speed 15 km s$^{-1}$. As the flare entered into the impulsive phase, the flux rope eruption accelerated to high linear speed of $\approx$ 110 km s$^{-1}$.}
	\label{Fig10}
\end{figure}

In order to investigate the evolution of the sigmoidal flux rope eruption, we plot a time-slice diagram for the duration 07:00 UT to 07:29 UT (Figure~\ref{Fig10}) which include both the precursor and impulsive phases of the flare. For the purpose, we have choosen a narrow slit S$_2$S$_1$ (Figure~\ref{Fig10}). Notably, the eruption of the leading front of the flux rope was observed to follow this path. The time-evolution of this slit is presented in Figure \ref{Fig10}b. For comparing the ascent of the flux rope with the evolutionary phases of the flare, we overplot the GOES SXR flux in 1--8 \AA~channel (shown by red curve in Figure~\ref{Fig10}b) which readily suggests that during the precursor phase of the flare \textit{i.e.} $\approx$ 07:00 UT--07:17 UT, the flux rope undergoes through a slow rise with a linear speed $\approx$ 15 km s$^{-1}$. After $\approx$ 07:17 UT \textit{i.e.} as the flare enters into the impulsive phase, the flux rope attains an acceleration phase with sudden rise in its speed to $\approx$ 110 km s$^{-1}$.

\subsection{Multi-wavelength Flare Evolution} 
      \label{S-flare evolution}    

\begin{figure}[H]    
	\centerline{\includegraphics[width=1.0\textwidth,clip=]{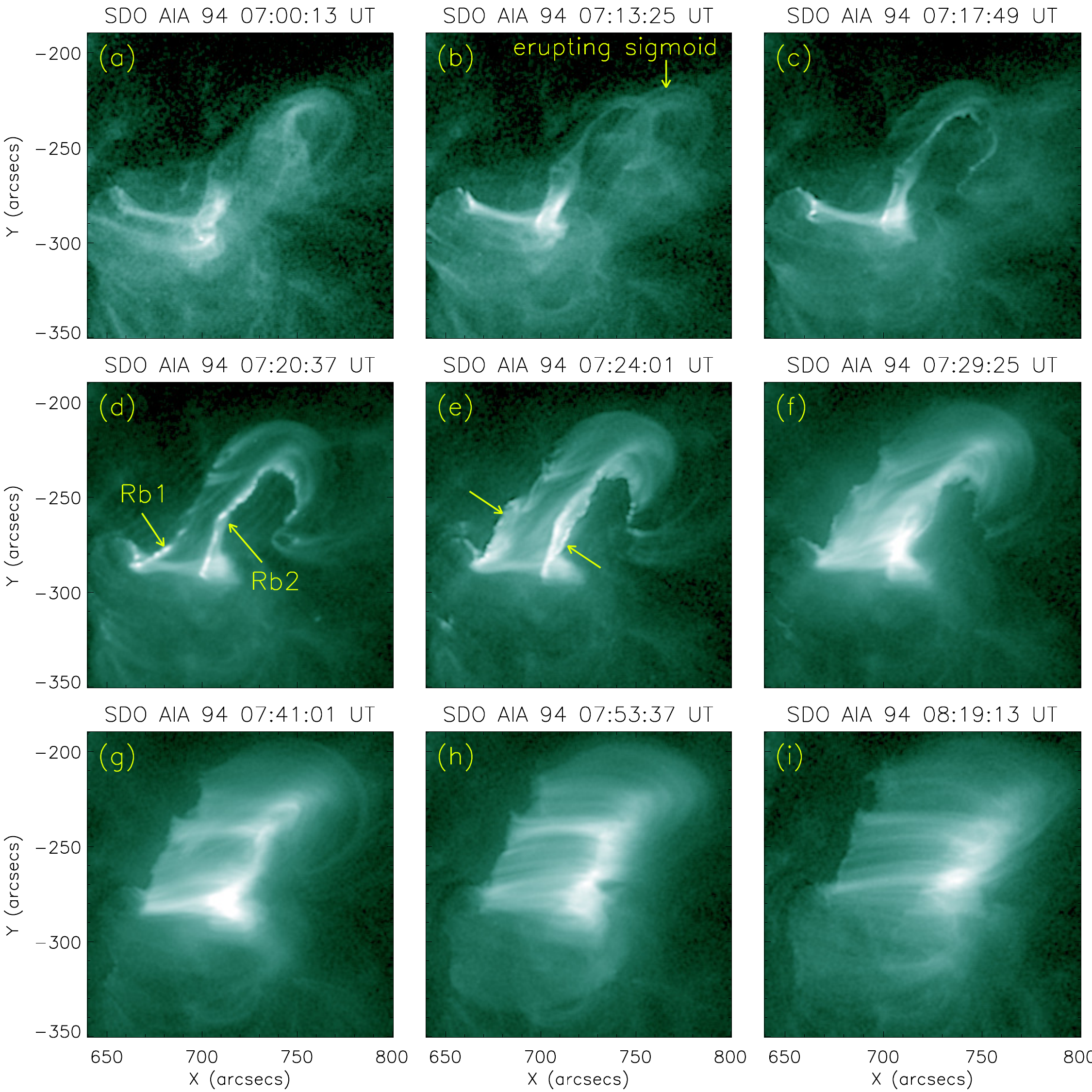}
	}
	\caption{Series of selected SDO/AIA 94 {\AA} images showing the evolution of the flare during the main phase. Arrows in panel (d) and (e) indicate the two ribbons of the M1.2 flare. An animation of this figure is provided in the supplementary materials.}
	\label{Fig7}
\end{figure}

In Figure \ref{Fig7}, we present a series of a few selected AIA 94 \AA\ images showing the evolution of the AR during different phases of the M1.2 flare. A clear erupting loop-like structure can be identified in Figure~\ref{Fig7}b. With the onset of the impulsive phase (i.e. from $\approx$ 07:17 UT), we observed a set of parallel ribbons (indicated as Rb1 and Rb2 in Figure~\ref{Fig7}d) which are situated on both sides of the PIL. With the evolution of the flare, the brightness of the ribbons enhances. Also, the separation between them increases. During the gradual phase, we observed formation of highly structured post-flare arcade encompassing the entire active region (Figures \ref{Fig7}f--i).

\begin{figure}[H]    
	\centerline{\includegraphics[width=1.0\textwidth,clip=]{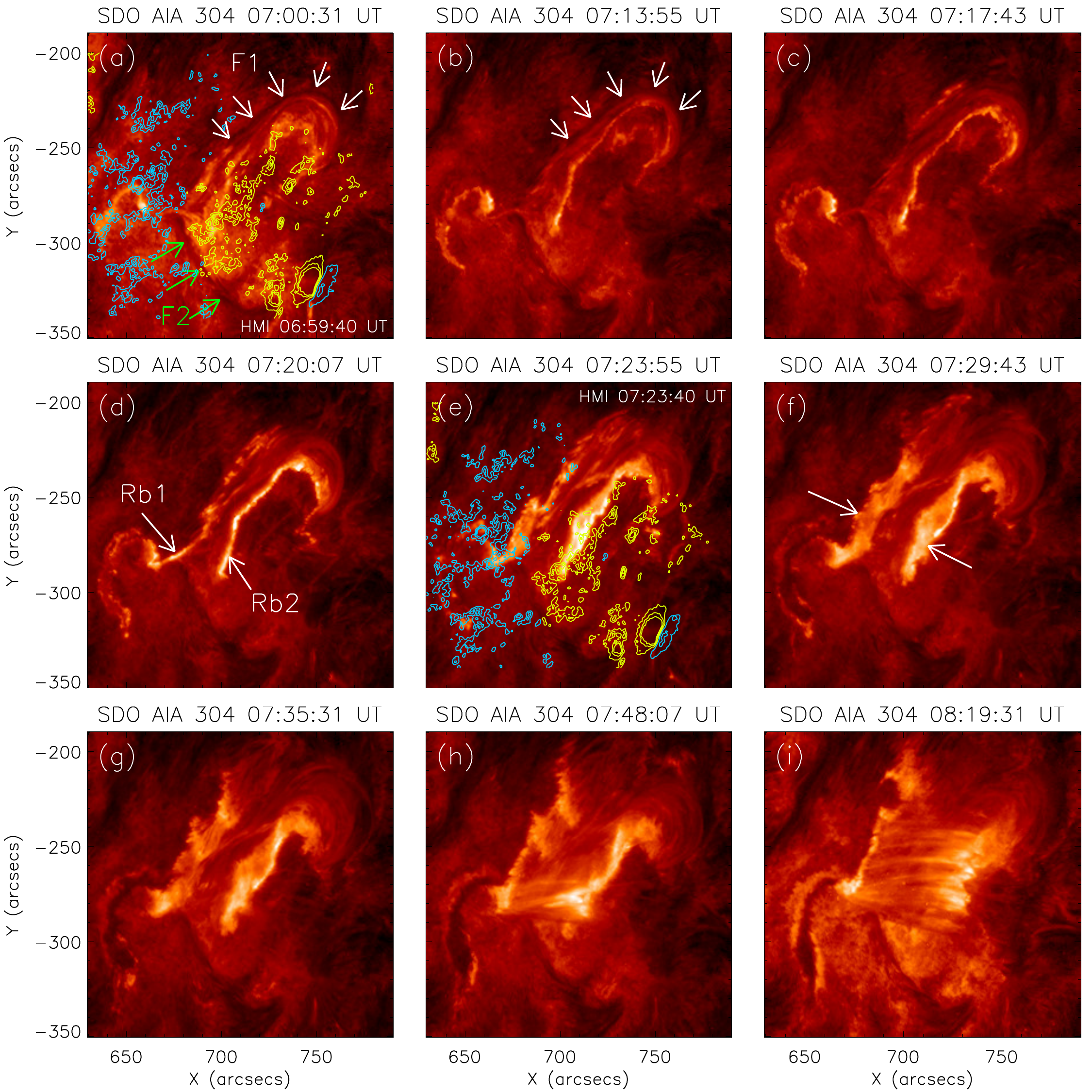}
	}
	\caption{Series of selected SDO/AIA 304 {\AA} images showing the evolution of the flare during the main phase. White and green arrows in panel (a) show the pre-existing filaments F1 and F2, respectively in the AR. The erupting filament F1 is shown by white arrows in panel (b). Panels (d-h) and (i) show two-ribbons (two white arrows in panels d and f) and post-flare loop arcade, respectively in the AR. HMI line-of-sight magnetogram is overplotted in panels (a) and (e). The positive and negative polarities are shown by yellow and blue contours, respectively, with contour levels set as $\pm$150, $\pm$400, $\pm$600 G. An animation of this figure is provided in the supplementary materials.}
	\label{Fig8} 
\end{figure}

The AIA 304~\AA\ images clearly show filaments in the precursor phase and their subsequent evolution during the M-class flare. In Figure~\ref{Fig8}, we show a series of AIA 304~\AA~images. To show the distribution of the positive and negative polarities of the photospheric flux, HMI LOS magnetogram is overplotted in Figures~\ref{Fig8}a and e. We noticed the existance of two filaments along the PIL: F1 in the northern part (white arrows in Figure~\ref{Fig8}a) and F2 in the southern part (green arrows in Figure~\ref{Fig8}a) of the AR. Filament F1 coincides with the ``elbow" of the sigmoidal structure implying a close spatial association with the sigmoid (\textit{cf.} Figures~\ref{Fig7} and \ref{Fig8}). Filament F2 is situated far from the sigmoid, but within the AR. It is noticed that the filament F1 activates and undergoes through eruption during the precursor phase (white arrows in Figure~\ref{Fig8}b). However, no significant changes are observed in the filament F2. The activation of the filament F1 starts at $\approx$ 07:13 UT and it successfully erupts by $\approx$ 07:20 UT. Afterward, we noticed the development of two-ribbon flare (Figure~\ref{Fig8}d). As discussed earlier, the increasing seperation between the ribbons can clealy be seen from Figure~\ref{Fig8}d--h (marked by white arrows in Figures~\ref{Fig8}d and f). The post-flare loop arcade start to appear from $\approx$ 08:19 UT (Figure~\ref{Fig8}i).

  \subsection{Radio observations} 
  
\begin{figure}[t!]    
  	\centerline{\includegraphics[width=1.0\textwidth,clip=]{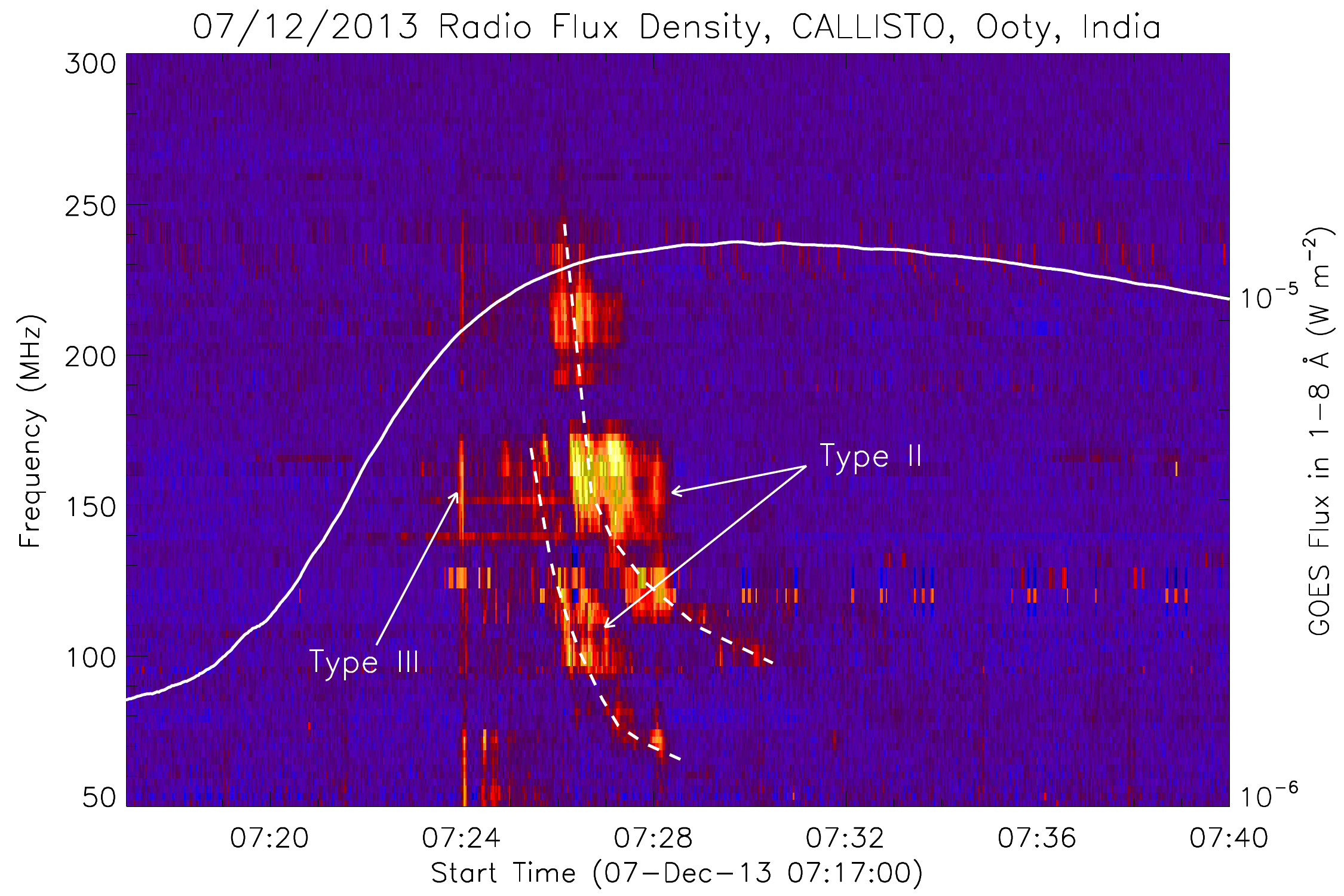}
  	}
  	\caption{Solar radio spectrum observed by CALLISTO spectrograph on 07 December 2013 from 07:17 UT to 07:40 UT within the frequency range of 50--300 MHz. Spetrograph shows a type III burst at $\approx$ 07:24 UT and a split-band type II burst between $\approx$ 07:26 and $\approx$ 07:29 UT. White solid curve over the spectrum shows the GOES SXR flux variation in 1--8 \AA\ channel. Two dash curves indicate the fundamental and harmonic bands of the type II radio burst.}
  	\label{Fig11}
  \end{figure}

  Figure~\ref{Fig11}~displays the radio dynamic spectrum obtained from the CALLISTO spectrograph within the frequency range of 50--300 MHz on 07 December 2013 during 07:17 UT--07:40 UT. For comparison, we overplot the GOES SXR flux in 1--8~\AA~wavelength channel within the same time axis (shown by white curve). From Figure \ref{Fig11}, we readily observe that during the impulsive phase of the flare, a type III solar radio burst occurs at $\approx$ 07:24 UT. This type III burst is followed by a split-band type II solar radio burst during $\approx$ 07:26 UT--07:29 UT. The fundamental band of type II burst is observed with starting frequency of $\approx$ 173 MHz. Using the observations of fundamental band, we have calculated the shock parameters using the Gopalswamy model \citep{Gopalswamy2013}. The shock speed is estimated as $\approx$ 1143 km s$^{-1}$ and the average height of shock formation is $\approx$ 1.32 R$_\odot$. The eruptive flare, however, lacked solar proton event.
  
  \subsection{CME in the near-Sun region} 
   
\begin{figure}   
   	\centerline{\includegraphics[width=1.0\textwidth,clip=]{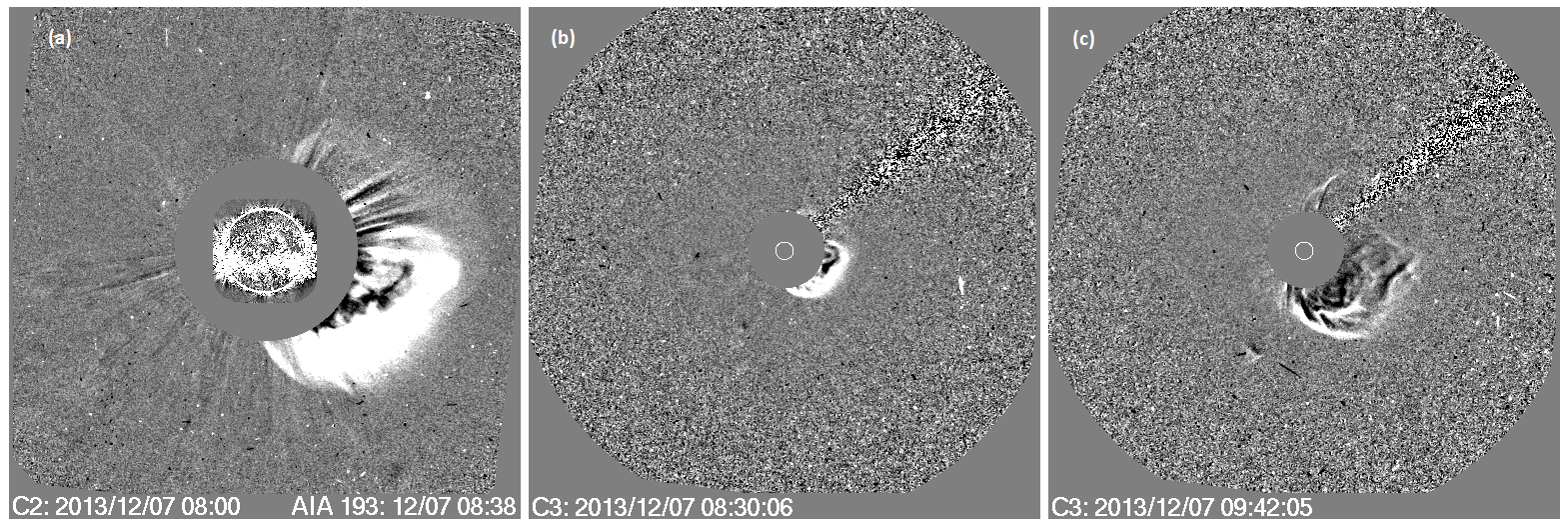}
   	}
   	\caption{Running difference images derived from SOHO/LASCO C2 (panel a) and C3 (panel b and c) coronagraph showing the propagation of halo CME originated from AR 11909 during the M1.2 flare on 07 December 2013.}
   	\label{Fig12}
   \end{figure}

\begin{figure}    
   	\centerline{\includegraphics[width=1.0\textwidth,clip=]{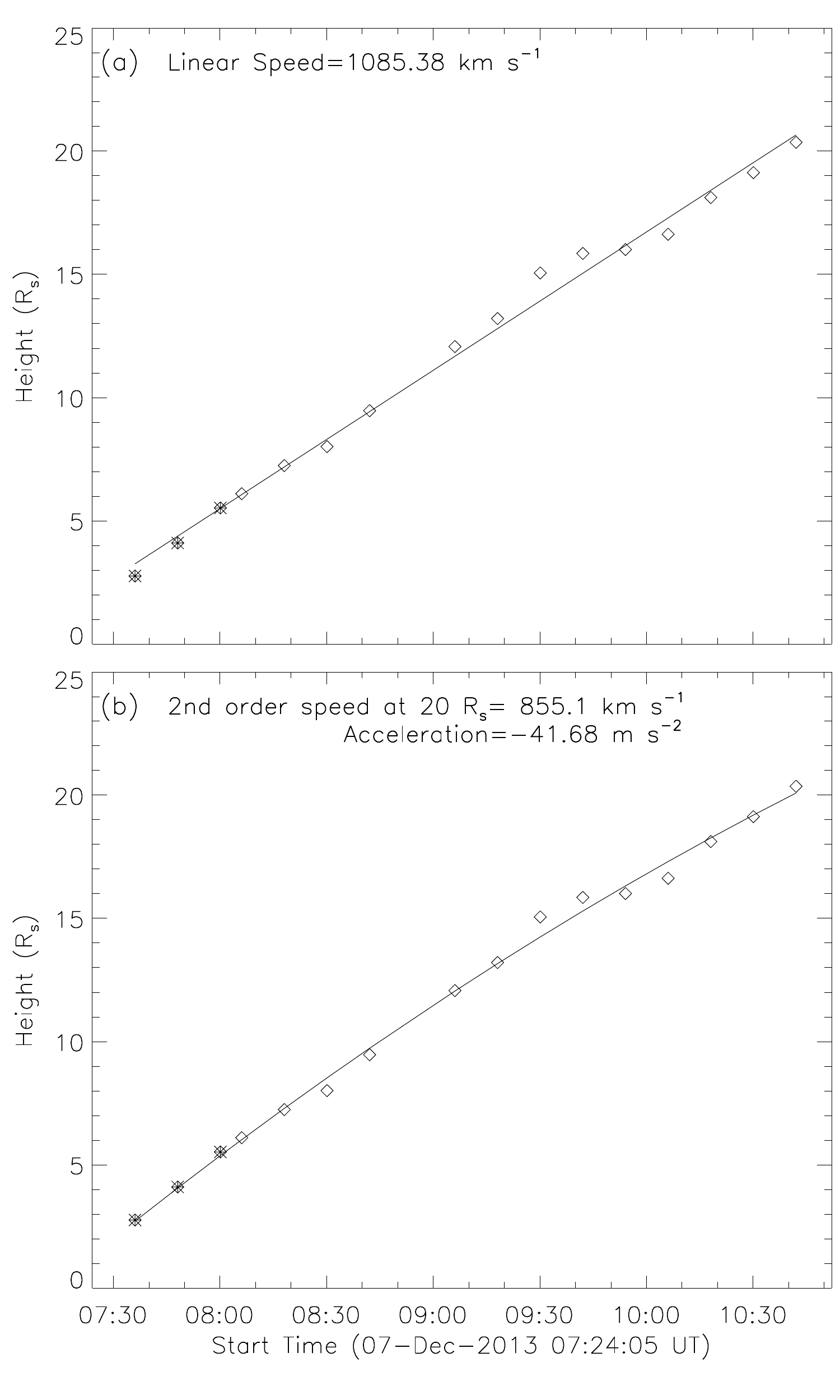}
   	}
   	\caption{Height-time plot of the CME on 07 December 2013 between 07:24 UT and 10:42 UT with linear (panel a) and second order (panel b) fitting. Asterisk (*) and diamond shaped symbols show the height of CME observed by the C2 and C3 coronagraphs, respectively, of SOHO/LASCO. Linear fit in panel (a) provides the linear speed of the CME to be $\approx$ 1085 km s$^{-1}$ while the second order fit in panel (b) shows that the CME is deccelerating with acceleration of $\approx$ 41.68 m s$^{-2}$.}
   	\label{Fig13}
   \end{figure}

   According to SOHO/LASCO CME catalogue, the flux rope eruption is associated with a halo CME which is detected by the LASCO C2 and C3 coronagraph (Figure~\ref{Fig12}). The height-time plot corresponding to the CME within the LASCO FOV is provided in Figure~\ref{Fig13}. The CME is first detected in the LASCO C2 coronagraph at 07:36 UT at a height of $\approx$ 2.76 R$_\odot$ and is visible in C2 till 08:00 UT (the C2 data points are denoted by asterisk symbol (*) in  Figures \ref{Fig13}a and b). The CME is observed till 10:42 UT by C3, when it reached the height of 20.36 R$_{\odot}$ (denoted by diamond symbols in Figures \ref{Fig13}a and b). By applying a linear fit to the height-time data points (Figure~\ref{Fig13}a), we calculate the linear speed of the CME to be $\approx$ 1085 km s$^{-1}$. By using second order fit to the height-time data points, the acceleration of the CME is estimated as $\approx$ -41.68 m s$^{-2}$. Thus, we can infer that the flux rope eruption led to a deccelerated high speed halo CME. The speed of the CME is calculated as $\approx$ 855 km s$^{-1}$ at a height of 20 R$_\odot$. 
   
\section{Discussion} 
      \label{discussion} 
      
We study the formation of a transient coronal sigmoid and its subsequent disruption which led to an M-class flare and fast halo CME. The activities occurred in the AR 11909 on 07 December 2013. We summarize the chronology of the event in Table 1.

We have observed the build-up and formation phase of a transient coronal sigmoid, an evidence of magnetic flux rope, in hot EUV channel of 94 \AA. The AIA 94 \AA\ observations cleary reveal that the formation process of coronal sigmoid witnessed the joining of two twisted loop-like structures with an overall enhancement in the brighteness which likely suggests that reconnection-coupling of the two flux ropes resulted into the build-up of a large sigmoidal flux rope. About $\approx$ 20 min after the complete formation of the coronal sigmoid, the eruptive activities started from the region which ultimately led to a two-ribbon flare and fast CME ($\approx$ 1085 km s$^{-1}$). The eruption of sigmoidal flux rope caused large-scale restructuring of the active region corona and typical ``sigmoid-to-arcade" transformation is observed. Transient sigmoids imply those sigmoids which appear for a shorter time duration and usually observed to form prior to an eruptive flare \citep[see \textit{e.g.}][]{Pevtsov1996, Bhuwan2017}. To understand the evolution of the sigmoidal flux rope in the AR, we have explored the coupling between different lower atmospheric layers \textit{i.e} photosphere, chromosphere, and transition region. The observations of the chromosphere and transition region in H$\alpha$ (Figure~\ref{Fig1}c) and AIA 304 \AA~channels (Figure~\ref{Fig8}) indicate the existence of two filaments along the PIL. Filament F1 coincides with the sigmoid. Due to the spatial correspondance of filament (F1) and sigmoid, we can infer that both features correspond to the same flux rope. Notably, filament F1 completely erupts with the eruptive-expansion of the sigmoidal flux rope. The existance of the filament has been widely accepted as evidence of the magnetic flux rope \citep[\textit{e.g.}][]{Bhuwan2017, Joshi2018SoPh, Mitra2020ApJ, Suraj2020}. Through the analysis of the HMI magnetogram, we observed the moving magnetic features along with magnetic field changes near the PIL. These features are supposed to play important roles in the formation of flux rope and its subsequent eruption \citep[\textit{e.g.}][]{Bhuwan2011ApJ, Upendra2014ApJ, Vasantharaju2019, Prabir2020SoPh}.

The photospheric observations of the AR also depicts the flux cancellation near the PIL. As a consequence, we found a decrease in the net flux of the AR. It is likely that the flux cancellation would lead to the tether-cutting reconnection \citep{Moore1992LNP, Moore2001} among opposite polarity foot points of the low-lying short loops that are rooted just across the PIL thereby facilitating the development of the flux ropes. With the progression of the build-up phase of the sigmoid, the flux rope acquires a large amount of magnetic free enery, which is released during the magnetic reconnection in the form of solar flares and CMEs \citep[\textit{e.g.}][]{Prabir2018ApJ, Prabir2020SoPh}. The AIA 94 \AA\ images suggest the coupling of two twisted loop-like structures during the build-up phase of the coronal sigmoid which correlates with our interpretation of observed flux cancellation process close to the PIL. Successful eruption of the flux rope is found to be followed by a two-ribbon solar flare of M1.2 class. The increasing separation between the flare ribbons is also noticed in the study. The flare ribbons represent the foot-points of the coronal loops involved in the coronal magnetic reconnection process \citep[\textit{e.g.} see,][]{Veronig2006AA, Bhuwan2007SoPh, BhuwanJ2009ApJ}. With the progression of a solar flare, the coronal loops rooted at the opposite polarity sides of the conjugate flare ribbons evolve from `stressed' configuration to nearly potential state \citep{Bhuwan2017ApJ85129}. 

The appearance of the sigmoidal structure in the AR is considered as a progenitor of CMEs in the source region \citep{Canfield2000}. The early evolution of the CMEs is associated with the activation of the flux rope. In this regard, we would like to further emphasize the two-phase evolution of the sigmoidal flux rope: slow rise and fast rise. The early dynamical evolution of the flux rope is identified by slow rise phase with a speed of $\approx$ 15 km s$^{-1}$ which points toward the initiation phase of the CME \citep{Zhang2001, Bhuwan2016ApJ, Prabir2019ApJ}. The evolution of the flux rope undergoes sudden transition from state of slow rise to fast rise with the drastic increase of its eruption speed to $\approx$ 110 km s$^{-1}$. The fast rise state essentially marks the impulsive acceleration phase of the CME \citep{Zhang2001}. Our measurement of the flux rope speed during its eruptive phase is in agreement with the speeds of plasmoids, reported mainly during the limb events \citep[\textit{e.g.}][]{Manoharan2003ApJ, Bhuwan2016ApJ}. Notably, the slow and fast rise phases of the sigmoidal flux rope are well correlated with the precursor and impulsive phases of the flare, respectively. In essence, the formation of transient sigmoid and activation of EUV flux rope support the scenario of the development of CME in the low corona \citep[see \textit{e.g.}][]{Mano2001ApJ, Cho2009ApJ, Joshi2018SoPh}

The radio dynamic spectrum shows the occurance of type III and type II radio bursts during fast rise phase of the flux rope or impulsive phase of the flare. Type III radio burst is observed at $\approx$ 07:24 UT and type II radio burst is observed during $\approx$ 07:26 UT--07:29 UT. Type III radio bursts are produced due to accelerated electrons propagating along the open magnetic field lines \citep[see review by][]{Reid2014RAA}. The rapidly accelerating flux rope structure would drive a coronal shock. The type II radio burst in the frequency range of $\approx$ 50--300 MHz implies the fast eruption of the CME-flux rope system in the lower coronal heights ($\approx$ 1.32 R$_\odot$). Considering the projected speed of erupting flux rope as $\approx$ 110 km s$^{-1}$ within the AIA 94 \AA\ FOV (see Figure \ref{Fig10}), the initiation time and formation height of type II radio burst correlate with the imaging observations. The CME is later detected on C2 coronagraph of LASCO at a height of 2.76 R$_\odot$ at 07:36 UT. Notably, the shock speed is found to be comparable to the speed of the CME in the LASCO FOV. 

\section{Conclusion}
      \label{conclusion} 
      
In this paper, we provide a multi-wavelength analysis of the build-up phase of a sigmoidal flux rope in the corona and its subsequent ``sigmoid-to-arcade" transformation. Notably, the sigmoid studied here is of transient category and its disruption started just $\approx$ 20 min after its complete formation. The coronal sigmoid is associated with a simple $\beta$-type AR which posseses relatively weaker and dispersed magnetic flux regions. Notably, both legs of the sigmoid are rooted in the magnetically dispersed regions, devoid of sunspots. The photospheric moving magnetic features and flux cancellation events are observed during the build-up phase of the sigmoid as well as its post-formation phase when the eruptive activities set-in. In particular, the observed flux cancellation near the PIL has important bearing toward the strengthening and subsequent activation of the magnetic flux rope by tether-cutting magnetic reconnections. These small-scale, episodic reconnection events will also contribute toward plasma heating which enhances the temperature of the newly formed flux rope. At this pre-eruption stage, the flux rope starts to become visible in SXR and hot EUV channels. The ``sigmoid-to-arcade" transformation accompanies a classical two-ribbon flare (of GOES M1.2 class) and a fast CME ($\approx$ 1085 km s$^{-1}$). The two-phase evolution of the flux rope clearly bifurcates the precursor and impulsive phases of the solar flare, which points toward the activation of flux rope by tether-cutting process and subsequent large-scale reconnection driven by expanding CME in the source region.

\begin{acks}
We would like to thank SDO team for their open data policy. SDO is NASA's mission under the Living with a Star (LWS) program. We also thank FHNW, Institute for Data Science in Brugg/Windisch, Switzerland for hosting the e-Callisto network. We also acknowledge National Astronomical Observatory of Japan (NAOJ), Mitaka, Japan for providing the H$\alpha$ images of the Sun. We are thankful to GOES for providing SXR data of the Sun. GOES is a joint mission of NASA and the
National Oceanic and Atmospheric Administration (NOAA). We acknowledge VAPOR which is a product of the National Center for Atmospheric Research's Computational and Information Systems Lab, supported by the U.S. National Science Foundation and by the Korea Institute of Science and Technology Information. Authors would like to thank anonymous referee for providing constructive comments and suggestions, which improved the presentation and scientific content of the article.  

\noindent 
{\bf Disclosure of Potential Conflict of Interest} The authors declare that they have no conflict of interest.
\end{acks}

  

\bibliographystyle{spr-mp-sola}






\end{article} 

\end{document}